\documentclass[aps,pra,epsfigure,onecolumn,preprint]{revtex4}
\usepackage{amsmath}
\usepackage{amsfonts}
\usepackage{epsfig}
\usepackage{epstopdf}

\newcommand{\miniket}[1]{\vert#1\rangle}

\newcommand{\minibra}[1]{\langle#1\vert}

\newcommand{\qed}{\nobreak \ifvmode \relax \else
      \ifdim\lastskip<1.5em \hskip-\lastskip
      \hskip1.5em plus0em minus0.5em \fi \nobreak
      \vrule height0.75em width0.5em depth0.25em\fi}

\begin{document}

\title{Measurement-induced generation of spatial entanglement in a two-dimensional quantum walk with single-qubit coin}

\author{C. Di Franco$^{1,2,3}$, M. Mc Gettrick$^4$, T. Machida$^5$ and Th. Busch$^1$}

\affiliation{$^1$ Department of Physics, University College Cork, Cork, Republic of Ireland\\
$^2$ Departamento de Qu\'imica F\'isica, Universidad del Pa\'is Vasco-Euskal Herriko Unibertsitatea, Apartado 644, E-48080 Bilbao, Spain\\
$^3$ Centre for Theoretical Atomic, Molecular, and Optical Physics, School of Mathematics and Physics, Queen's University, Belfast BT7 1NN, United Kingdom\\
$^4$ The De Br\'un Centre for Computational Algebra, School of Mathematics, The National University of Ireland, Galway, Republic of Ireland\\
$^5$ Meiji Institute for Advanced Study of Mathematical Sciences, Meiji University, 1-1-1 Higashimita, Tamaku, Kawasaki 214-8571, Japan}

\begin{abstract}
One of the proposals for the exploitation of two-dimensional quantum walks has been the efficient generation of entanglement. Unfortunately, the technological effort required for the experimental realization of standard two-dimensional quantum walks is significantly demanding. In this respect, an alternative scheme with less challenging conditions has been recently studied, particularly in terms of spatial-entanglement generation [C. Di Franco, M. Mc Gettrick, and Th. Busch, Phys. Rev. Lett. {\bf 106}, 080502 (2011)]. Here, we extend the investigation to a scenario where a measurement is performed on the coin degree of freedom after the evolution, allowing a further comparison with the standard two-dimensional Grover walk.
\end{abstract}

\maketitle

Quantum walks have raised considerable attention in the physics community due to their potential use in the design of new quantum algorithms~\cite{Aharonov:93,Kempe:03,venegas,santha,ambainis,konno,venegas2}. Moreover, two-dimensional quantum walks have been proposed for generating entanglement efficiently~\cite{QWEntanglementBose,QWEntanglementGoya}, a critical requirement for quantum metrology and quantum information processing~\cite{entanglementgeneral}. And yet, the experimental realization of known two-dimensional quantum walks is extremely challenging, due to the significant technological effort required. One of the crucial difficulties is embodied by the fact that higher-dimensional quantum coins have to be used. A simplification in this respect (where only a single-qubit coin is required) has been recently proposed~\cite{ourwalk}, and a comparison in terms of entanglement generation with respect to a well-known standard two-dimensional quantum walk (Grover walk) has been reported~\cite{longerversion}. However, the investigation has only considered the scenario where the coin degree of freedom is ignored after the evolution. Performing a coin measurement could actually represent an advantage in terms of entanglement generation: both the single-qubit case as well as the standard Grover quantum walk have to be explored. The study should also include an optimization process to find the best basis in which the coin has to be measured. This is precisely the scenario investigated in this paper.

For the sake of clarity, let us first briefly introduce the characteristics of the two quantum walks studied here. In both cases, as in general in any discrete-time quantum walk, the total state of the system under investigation is a vector in the composite Hilbert space ${\cal H}={\cal H}_W\otimes{\cal H}_C$. ${\cal H}_W$ (walker space) is an infinite-dimensional Hilbert space spanned by $\{\miniket{x,y}\}$, with $x$ and $y$ assuming all possible integer values. The coin space, as already stated, is different in the two cases, so we denote the coin space of the single-qubit quantum walk (from now on, {\it alternate quantum walk}) as ${\cal H}_C$ and the one of the Grover walk as ${\cal H}_{C'}$. ${\cal H}_C$ is a two-dimensional Hilbert space spanned by $\{\miniket{0},\miniket{1}\}$, while ${\cal H}_{C'}$ is four-dimensional, spanned by $\{\miniket{0},\miniket{1},\miniket{2},\miniket{3}\}$. Let us define the basis states of the total space ${\cal H}$ as $\{\miniket{x,y,c}\}$ and $\{\miniket{x,y,c'}\}$, respectively. Here, for the sake of simplicity, we use the notation $\miniket{x,y,c}=\miniket{x,y}_W\otimes\miniket{c}_C$ and $\miniket{x,y,c'}=\miniket{x,y}_W\otimes\miniket{c'}_{C'}$.

The evolution of the system in a discrete-time quantum walk is usually defined by a sequence of conditional shift and coin operations. In the case of the alternate quantum walk, we have two different conditional shift operations
\begin{equation}
\hat{S}_x=\sum_{i,j\in \mathbb{Z}}(\miniket{i-1,j,0}\minibra{i,j,0}+\miniket{i+1,j,1}\minibra{i,j,1})
\end{equation}
and
\begin{equation}
\hat{S}_y=\sum_{i,j\in \mathbb{Z}}(\miniket{i,j-1,0}\minibra{i,j,0}+\miniket{i,j+1,1}\minibra{i,j,1}),
\end{equation}
while the coin operation is the Hadamard gate
\begin{equation}
\hat{H}=\frac{1}{\sqrt{2}}
\begin{pmatrix}
1&1\\
1&-1
\end{pmatrix}\;.
\end{equation}
If we consider the walker component $\miniket{i,j}_W$ as describing the quantized position of the walker in the $x$ and $y$ directions with increasing numbers from left to right and from bottom to top, respectively, the effect of $\hat{S}_x$ is to move the walker one step to the left (right) when the coin component is in the state $\miniket{0}_C$ ($\miniket{1}_C$) and the effect of $\hat{S}_y$ is to move the walker one step down (up) when the coin component is in the state $\miniket{0}_C$ ($\miniket{1}_C$). In the alternate quantum walk, a single time step consists of two Hadamard operations and one movement each on the $x$ and $y$ directions, according to the following sequence: coin operation - movement on $x$ - coin operation - movement on $y$.

The shift operator for the Grover walk is 
\begin{equation}
\begin{split}
\hat{S}_G=\sum_{i,j\in \mathbb{Z}}(\miniket{i-1,j-1,0}\minibra{i,j,0}+\miniket{i-1,j+1,1}\minibra{i,j,1}\\
+\miniket{i+1,j-1,2}\minibra{i,j,2}+\miniket{i+1,j+1,3}\minibra{i,j,3})
\end{split}
\end{equation}
and the corresponding Grover coin operation is given by
\begin{equation}
\hat{G}=
\frac{1}{2}\left(
\begin{array}{rrrr}
-1&1&1&1\\
1&-1&1&1\\
1&1&-1&1\\
1&1&1&-1
\end{array}\right).
\end{equation}
The states of the computational basis of the coin $\miniket{0}_{C'}$, $\miniket{1}_{C'}$, $\miniket{2}_{C'}$ and $\miniket{3}_{C'}$ correspond, under the action of the shift operator, to movements in the left-down, left-up, right-down and right-up directions, respectively. A single time step consists here of a Grover coin operation and a movement on the $x$-$y$ plane according to $\hat{S}_G$.

In Ref.~\cite{ourwalk}, a comparison between the $x$-$y$ spatial-entanglement generation in the two different walks has been presented. Let us shortly comment on this particular kind of entanglement. Suppose that we want to entangle, in the schemes aforementioned, the orthogonal directions of the lattice on which the walker is moving. Clearly, this can also be seen from a different point of view, that is formally equivalent. We could have two particles both moving on a line (or even on two different lines). They do not interact directly, but they share a common degree of freedom, embodied by the coin. The whole formalism of the quantum walks presented so far can therefore be easily mapped to this composite system, by considering the one-dimensional movement of the first particle as the shift of the original walker on the $x$ direction, and the one-dimensional movement of the second particle as the shift of the original walker on the $y$ direction. The $x$-$y$ spatial entanglement corresponds, in this scenario, to the entanglement between the position of these two different particles.

In order to evaluate the $x$-$y$ spatial entanglement, in Ref.~\cite{ourwalk}, the degree of freedom embodied by the coin has been traced out. This corresponds to ignoring the coin state after the evolution. However, performing a coin measurement could actually represent an advantage in terms of entanglement generation. The investigation of this {\it measurement-induced} entanglement, seen from the two-particles point of view, is strongly related to the study performed in Ref.~\cite{QWEntanglementBose}. The difference between the analysis presented there and the alternate-walk case here is that, in that paper, no coin operation is performed between the movements of the two particles (in our case, a Hadamard gate is applied). Moreover, here we are interested in the average entanglement generated, so we sum the amount of the entanglement corresponding to each result of the measurement multiplied by the probability to obtain that result.

\begin{figure}[t]
\psfig{figure=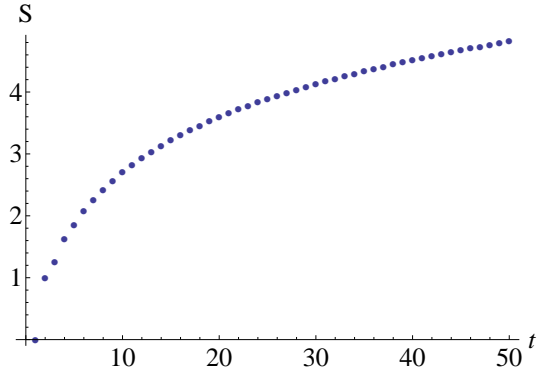,width=7cm}
\caption{Measurement-induced $x$-$y$ spatial entanglement (in terms of the Von Neumann entropy $S$ of the reduced density matrices) against the number of time steps $t$, for the alternate quantum walk with the initial state of the coin as in Eq.~(\ref{eq:AlternateInitial}) and the measurement performed on the computational basis $\{\miniket{0},\miniket{1}\}$.}
\label{fig:entanglementalternate}
\end{figure}

Let us start the investigation with the alternate quantum walk. The first attempt could be, following the lines of Ref.~\cite{QWEntanglementBose}, to consider the coin measurement on the computational basis $\{\miniket{0},\miniket{1}\}$. Clearly, after the measurement, the walker state is pure (the total state of the system, before the measurement, is pure, due to the unitarity of the evolution). It is thus possible to use the von Neumann entropy of the reduced density matrices to evaluate the measurement-induced entanglement. For this, we need to calculate $S(\rho_x)=S(\rho_y)$, where $S(\rho)=-Tr(\rho\log_2\rho)$ and $\rho_x$ ($\rho_y$) is the reduced density matrix of the $x$-position ($y$-position). The results for a number of time steps up to $t=50$ are shown in Fig.~\ref{fig:entanglementalternate}.
The initial state of the coin has been chosen, in this case, as
\begin{equation}
\frac{1}{\sqrt{2}}(\miniket{0}_C+i\miniket{1}_C),
\label{eq:AlternateInitial}
\end{equation}
which gives a symmetric evolution of the spatial probability distribution for the alternate quantum walk with respect to the $x$ and $y$ axes~\cite{longerversion}. It is easy to check that any initial condition of the form
\begin{equation}
\frac{1}{\sqrt{2}}(\miniket{0}_C+e^{i\alpha}\miniket{1}_C),
\end{equation}
with $\alpha\in[0,2\pi]$ generates the same amount of entanglement (notice that all these states have their Bloch vector~\cite{nielsen} orthogonal to those of the states in the measurement basis).

\begin{figure}[t]
\psfig{figure=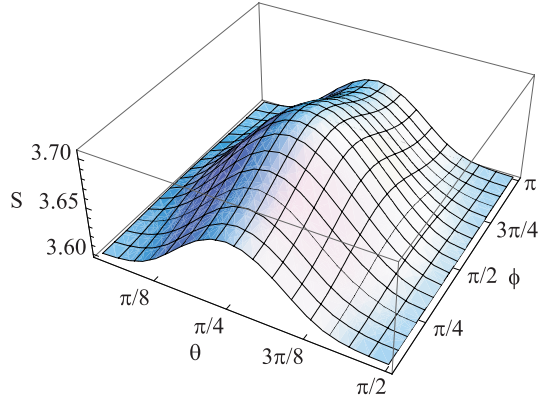,width=7cm}
\caption{Measurement-induced $x$-$y$ spatial entanglement (in terms of the Von Neumann entropy $S$ of the reduced density matrices) for the alternate quantum walk with the initial state of the coin as in Eq.~(\ref{eq:AlternateInitial}), where the projection is performed on the state in Eq.~(\ref{eq:measurementbasis}) and its orthogonal, against $\theta$ and $\phi$. The number of time steps considered here is $t=20$.}
\label{fig:entanglementalternaterotated}
\end{figure}

\begin{figure}[t]
\psfig{figure=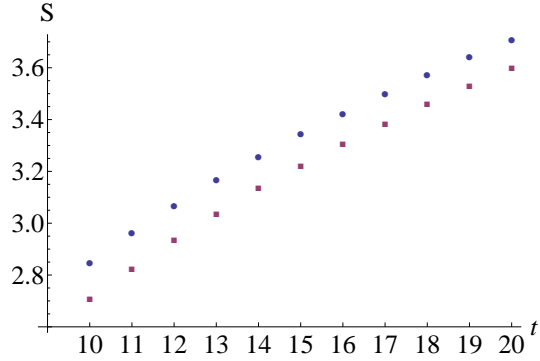,width=7cm}
\caption{Measurement-induced $x$-$y$ spatial entanglement (in terms of the Von Neumann entropy $S$ of the reduced density matrices) against the number of time steps $t$, for the alternate quantum walk with the initial state of the coin as in Eq.~(\ref{eq:AlternateInitial}) and the measurement performed on the state in Eq.~(\ref{eq:measurementbasis}) and its orthogonal, with $\theta=\pi/4$ and $\phi=\pi/2$ (blue circles). For comparison, the entanglement generated when the measurement is performed on the computational basis is also shown (purple squares).}
\label{fig:entanglementalternatemax}
\end{figure}

Clearly, one can choose a different basis on which the coin has to be projected during the measurement stage. In order to find the basis that maximizes the average entanglement generated, let us consider the two generic states of this basis as
\begin{equation}
\cos\theta\miniket{0}+e^{i\phi}\sin\theta\miniket{1},
\label{eq:measurementbasis}
\end{equation}
and the orthogonal one, with $\theta\in[0,\pi/2]$ and $\phi\in[0,\pi]$. We have studied the generated entanglement against $\theta$ and $\phi$ for different values of $t$ [fixing the initial condition of the coin as in Eq.~(\ref{eq:AlternateInitial})] and found that, for $t>6$, the maximum always corresponds to $\theta=\pi/4$ and $\phi=\pi/2$, with the minimum corresponding to the measurement on the computational basis (as a significant example, we present in Fig.~\ref{fig:entanglementalternaterotated} the plot obtained for $t=20$).
This allows us to evaluate the maximum of the generated entanglement against $t$, by fixing the optimal measurement basis. The results for a number of time steps between $t=10$ and $t=20$ are shown in Fig.~\ref{fig:entanglementalternatemax}.

\begin{figure}[t]
\psfig{figure=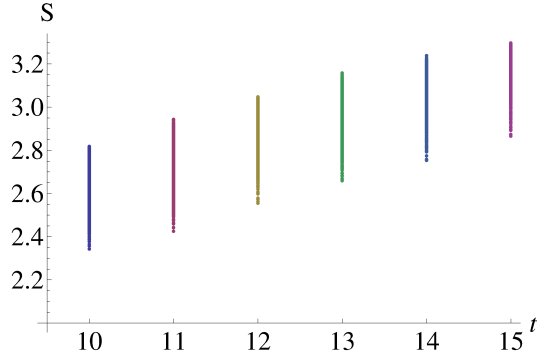,width=7cm}
\caption{Measurement-induced $x$-$y$ spatial entanglement (in terms of the Von Neumann entropy $S$ of the reduced density matrices) against the number of time steps $t$, for the Grover walk with the initial state of the coin as in Eq.~(\ref{eq:GroverInitial}) and the measurement performed on an ensamble of $5\cdot 10^3$ randomly generated bases.}
\label{fig:entanglementgroverrandom}
\end{figure}

Let us now consider the entanglement generated in the Grover walk. Also in this case, we have to find the maximizing basis. Following the investigation performed for the alternate quantum walk, we start from the computational basis $\{\miniket{0},\miniket{1},\miniket{2},\miniket{3}\}$. The initial state of the coin has been chosen, here, as
\begin{equation}
\frac{1}{2}(\miniket{0}_{C'}-\miniket{1}_{C'}-\miniket{2}_{C'}+\miniket{3}_{C'}),
\label{eq:GroverInitial}
\end{equation}
which corresponds to the non-localized case of the Grover walk~\cite{longerversion}. Interestingly, we obtain the same values presented in Fig.~\ref{fig:entanglementalternate}. This numerical analysis leads us to claim that {\it performing a coin measurement on the corresponding computational basis for the alternate quantum walk and the Grover walk after the same number of time steps generates exactly the same amount of $x$-$y$ spatial entanglement}. The next step is clearly, also in this case, to find the basis maximizing the generation of measurement-induced entanglement. For the Grover walk, the finding of the optimal measurement basis is more demanding, due to the larger coin space (and therefore the larger number of variables in the definition of the basis states, over which we should maximize). We have thus used a different method. We have randomly generated several bases and calculated the entanglement corresponding to measurements on them. For instance, we present in Fig.~\ref{fig:entanglementgroverrandom} the results obtained in this way for a number of time steps between $t=10$ and $t=15$ (we have generated $5\cdot 10^3$ different bases).
We have then studied the entanglement generated with the particular basis
\begin{equation}
\{\frac{\miniket{0}+\miniket{3}}{\sqrt{2}},\frac{\miniket{0}-\miniket{3}}{\sqrt{2}},\frac{\miniket{1}+\miniket{2}}{\sqrt{2}},\frac{\miniket{1}-\miniket{2}}{\sqrt{2}}\}
\label{eq:maxbasis}
\end{equation}
and found that this always corresponds to the top points in the previous plot. For the sake of completeness, a basis giving values of entanglement corresponding to the bottom points in the previous plot is
\begin{equation}
\{\frac{\miniket{0}+\miniket{1}}{\sqrt{2}},\frac{\miniket{0}-\miniket{1}}{\sqrt{2}},\frac{\miniket{2}+\miniket{3}}{\sqrt{2}},\frac{\miniket{2}-\miniket{3}}{\sqrt{2}}\}.
\end{equation}
We can therefore use the basis in Eq.~(\ref{eq:maxbasis}) to evaluate the maximum entanglement generated against $t$. Again, we have found that the values are exactly the same as those presented in Fig.~\ref{fig:entanglementalternatemax}. This numerical analysis leads us to claim that also {\it performing a coin measurement on the proper maximizing basis for the alternate quantum walk and the Grover walk after the same number of time steps generates exactly the same amount of $x$-$y$ spatial entanglement}.

We have studied the generation of entanglement obtained when a measurement is performed on the coin after the evolution of the alternate quantum walk as well as the Grover walk. We have shown that, properly choosing the measurement basis, in both the cases the entanglement can be increased with respect to the value obtained with the measurement on the computational basis. We have presented the optimal basis for each walk and noticed that the maximum of entanglement generated by measuring on these bases is the same for both the walks. This means that, even if after the evolution the amounts of coin-position and $x$-$y$ spatial entanglements are different in the alternate quantum walk and the Grover walk~\cite{ourwalk}, proper measurements are able to {\it localize}~\cite{verstraete} the same amount of spatial entanglement at the end of the two processes.

\section{Acknowledgments}
We thank C. Gillis for discussions. We acknowledge support from Science Foundation Ireland under Grants No. 05/IN/I852 and No. 10/IN.1/I2979. C.D.F. was supported by the Irish Research Council for Science, Engineering and Technology, the Basque Government grant IT472-10, and the UK EPSRC grant EP/G004579/1 under the ``New directions for EPSRC research leaders" initiative.


\begin{thebibliography}{000}

\bibitem{Aharonov:93} Y. Aharonov, L. Davidovich, and N. Zagury, Phys. Rev. A {\bf 48}, 1687 (1993).

\bibitem{Kempe:03} J. Kempe, Contemp. Phys. {\bf 44}, 307327 (2003).

\bibitem{venegas} S. E. Venegas-Andraca, {\it Quantum Walks for Computer Scientists} (Morgan and Claypool, San Rafael, CA, 2008).

\bibitem{santha} M. Santha, Lect. Notes Comput. Sci. {\bf 4978}, 31 (2008).

\bibitem{ambainis} A. Ambainis, Lect. Notes Comput. Sci. {\bf 4910}, 1 (2008).

\bibitem{konno} N. Konno, in {\it Quantum Potential Theory, Lecture Notes in Mathematics} {\bf 1954}, 309, edited by U. Franz and M. Sch\"urmann (Springer-Verlag, Heidelberg, 2008).

\bibitem{venegas2} S. E. Venegas-Andraca, Quantum Inf. Process. {\bf 11}, 1015 (2012).

\bibitem{QWEntanglementBose} S. E. Venegas-Andraca and S. Bose, arXiv:0901.3946 (2009).

\bibitem{QWEntanglementGoya} S. K. Goya and C. M. Chandrashekar, J. Phys. A {\bf 43}, 235303 (2010).

\bibitem{entanglementgeneral} R. Horodecki {\it et al.}, Rev. Mod. Phys. {\bf 81}, 865 (2009).

\bibitem{ourwalk} C. Di Franco, M. Mc Gettrick, and Th. Busch, Phys. Rev. Lett. {\bf 106}, 080502 (2011).

\bibitem{longerversion} C. Di Franco, M. Mc Gettrick, T. Machida, and Th. Busch, Phys. Rev. A {\bf 84}, 042337 (2011).

\bibitem{nielsen} M. A. Nielsen and I. L. Chuang, {\it Quantum Computation and Quantum Information} (Cambridge University Press, Cambridge, United Kingdom, 2000).

\bibitem{verstraete} F. Verstraete, M. Popp, and J. I. Cirac, Phys. Rev. Lett. {\bf 92}, 027901 (2004).

\end{thebibliography}
\end{document}